\documentclass[prb,twocolumn,amsmath,amssymb,superscriptaddress]{revtex4-1}
\usepackage[utf8x]{inputenc}
\usepackage{graphicx}
\usepackage{dcolumn}
\usepackage{bm}
\usepackage{color}

\begin{document}

\preprint{APS/123-QED}

\title{Interaction driven polarization shift in the $t-V-V'$ lattice
  fermion model at half filling: emergent Haldane phase}

\author{Bal\'azs Het\'enyi} \affiliation{Department of Theoretical
  Physics and MTA-BME ``Momentum'' Topology and Correlation Research
  Group, Budapest University of Technology and Economics, 1521
  Budapest, Hungary} \affiliation{Department of Physics, Bilkent
  University, TR-06800 Bilkent, Ankara, Turkey}

\begin{abstract}
We study the $t-V-V'$ model in one dimension at half-filling.  It is
known that for large enough $V$ fixed, as $V'$ is varied, the system
goes from a charge-density wave into a Luttinger liquid, then a
bond-order, and then a second charge density wave phase.  We find that
the Luttinger liquid state is further split into two, separating parts
with distinct values of the many-body polarization Berry phase.
Inside this phase, the variance of the polarization is infinite in the
thermodynamic limit, meaning that even if the polarization differs, it
would not be measurable.  However, in the gapped phases on each side
of the Luttinger liquid, the polarization takes a different measurable
value, implying topologically distinction.  The key difference is that
the large-$V'$ phases are link-inversion symmetric, while the
small-$V'$ one is site-inversion symmetric.  We show that large-$V'$
phase can be related to an $S=1$ spin chain, and exhibits many
features of the Haldane phase.  The lowest lying states of the
entanglement spectrum display different degeneracies in the two cases,
and we also find string order in the large-$V'$ phase.  We also study
the system under open boundary conditions, and suggest that the number
of defects is related to the topology.
\end{abstract}

\maketitle

\section{Introduction}

Topological condensed matter systems constitute an active research
area.  Topological band insulators are well
understood.~\cite{Bernevig13,Asboth16,Franz13} Quantum phase
transitions occur when the relevant topological invariant
($\mathbb{Z}$ or $\mathbb{Z}_2$) undergoes a finite change at a gap
closure {\it point}.  These systems also obey the bulk-boundary
correspondence principle (BBCP), which predicts the existence of edge
states in the topologically non-trivial phases.

Recently, attention has
focused~\cite{Franz13,Raghu08,Fidkowski10,Pollmann10,Fidkowski11,Turner11,Pollmann12,Manmana12,Yoshida14,Rachel18}
on interacting systems.  An early result is the Haldane
conjecture,~\cite{Haldane83a,Haldane83b,Affleck89} which is based on a
field-theoretical mapping of the Heisenberg model to a continuum one,
and states that $S=1$ spin chains are topologically non-trivial and
exhibit spin-$\frac{1}{2}$ edge states.  A useful scheme to visualize
this state of affairs is the Affleck, Kennedy, Lieb, and Tasaki (AKLT)
wave function, also known as the valence bond solid (VBS), a model for
$S=1$ systems.  Recently Oshikawa~\cite{Oshikawa92} extended the AKLT
wave function to arbitrary integer spin models.  Pollmann et
al.~\cite{Pollmann12} showed that topological protection is present
only in odd-$S$ systems, and the protecting symmetries are
time-reversal, dihedral rotation, and link inversion.

\begin{figure}[ht]
 \centering
 \includegraphics[width=\linewidth,keepaspectratio=true]{pd_newn.eps}
 \includegraphics[width=\linewidth,keepaspectratio=true]{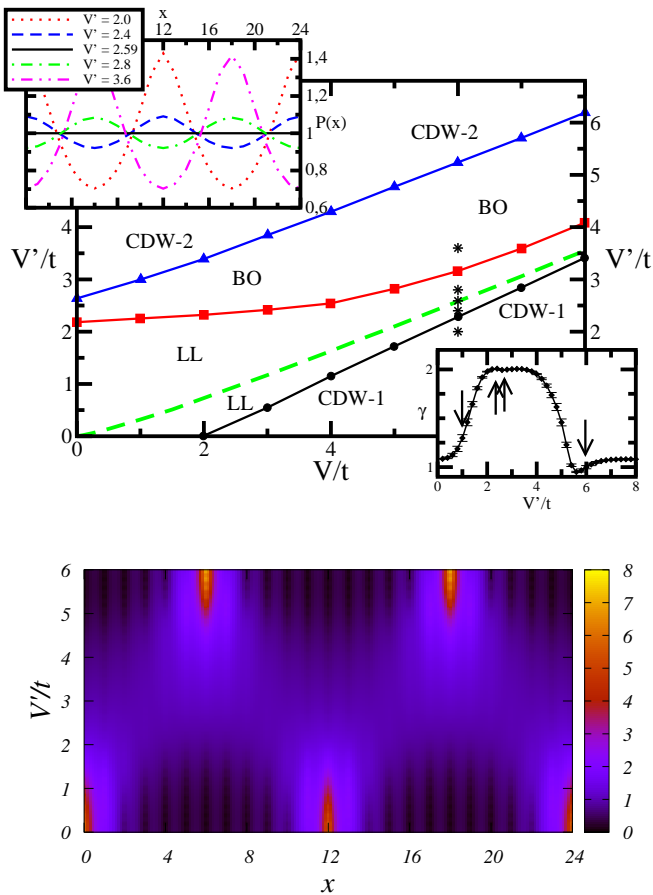}
 \caption{Upper panel: Phase diagram of the $t-V-V'$ model at
   half-filling.  The phase lines separating the charge density wave
   (CDW), Luttinger liquid (LL), bond-order (BO), and second charge
   density wave (CDW-2) phases were determined by Mishra {\it et al.}
   (Ref. \onlinecite{Mishra11}).  The thick dashed line inside the LL
   phase indicates the main finding of this paper, where the
   polarization undergoes a discrete change.  Along this line the
   polarization distribution is flat.  The maximum of the polarization
   shifts on either side.  Upper panel, upper left inset: polarization
   distribution for systems defined by stars in the main figure of the
   upper panel, $V=6;V'=2.0, 2.4, 2.59, 2.8, 3.6$.  These points are
   indicated with asterisks in the main figure.  The points are in the
   phases CDW, LL (below polarization switch), LL (where polarization
   switch occurs), LL (above polarization switch), BO, respectively.
   Exact diagonalization calculations with periodic boundary
   conditions.  \textcolor{red}{Upper panel, lower right inset: size
     scaling exponent of the variance of the polarization.  Arrows
     indicate the four cases shown in Fig. \ref{fig:finite_tvvp}.}
   Lower panel: heat map of the polarization distribution $P(x)$ as a
   function of $V'/t$, $V=6.0$.  Exact diagonalization calculations
   with periodic boundary conditions.  }
 \label{fig:pd_tvvp}
\end{figure}

The generalization of the idea of a topological invariant to the
many-body case is also a crucial question, since topological
invariants~\cite{Bernevig13,Asboth16,Franz13} in non-interacting
systems are integrals over Bloch states.  Manmana et
al.~\cite{Manmana12} define an invariant using the single particle
Green's function and the chiral symmetry operator.  The invariant
obtained this way reduces to the known invariant if the system is
non-interacting.  In the interacting case, topological edge states can
arise in three ways: poles or zeros in the Green's function
(single-particle effects) or spontaneous symmetry breaking at the edge
(many-body effect).  The latter is not necessarily picked up by a
topological invariant defined based on the single-particle Green's
function.  The polarization Berry phase~\cite{Resta98,Resta99} reduces
to the Zak phase when a non-interacting system is considered, however,
in the interacting case it is a genuine many-body expectation value.

In this paper we study the one-dimensional $t-V-V'$ interacting
lattice model of spinless fermions.  $t$ denotes the hopping
parameter, $V$ the nearest neighbor interaction, and $V'$ the next
nearest neighbor interaction.  It is known~\cite{Mishra11} that at
large enough $V$ a scan in $V'$ will find four phases: charge density
wave (CDW-1), Luttinger liquid (LL), bond-order (BO), and a different
charge density wave (CDW-2) phase.  Our central finding is that at a
critical $V_c'$ the LL phase is split into two parts.  For $V'<V_c'$
the Berry phase is zero, for $V'>V_c'$ it is $\pi$.  In the LL phase,
in the thermodynamic limit, the variance of the polarization diverges
with system size, thus the different polarization averages are not
measurable (expected for a gapless phase).  However, on the different
sides of the LL phase, the phases are such that the polarizations are
measurable, and the discrete difference between the two
implies~\cite{Watanabe18} topological distinction.  In
particular we find that the CDW-2 exhibits parallels to a Haldane
phase.~\cite{Haldane83a,Haldane83b} We show this via a mapping of our
original Hamiltonian to an $S=1$ spin model, by calculating the
entanglement spectrum, and by showing that hidden antiferromagnetic
(HAFM) order as well as finite range string correlation, as defined by
den Nijs and Rommelse~\cite{denNijs89} is present.  We also analyze
the system with open boundary conditions: our results here suggest
that the number of defects in a particular ordered phase may be
connected to the value of the topological invariant.

The paper is organized as follows.  In section \ref{sec:Model_method}
the $t-V-V'$ model is presented, as well as its connection to
integer-$S$ quantum spin chains.  In section \ref{sec:polamp} the
polarization amplitude is introduced.  It is shown that link inversion
gives rise to a non-trivial Berry phase, and several variants of the
Lieb-Schultz-Mattis~\cite{Lieb61} (LSM) theorem pertinent to our study are
derived.  In section \ref{sec:ED} our numerical results are presented,
in section \ref{sec:con} we conclude our work.

\begin{figure}[ht]
 \centering
 \includegraphics[width=\linewidth,keepaspectratio=true]{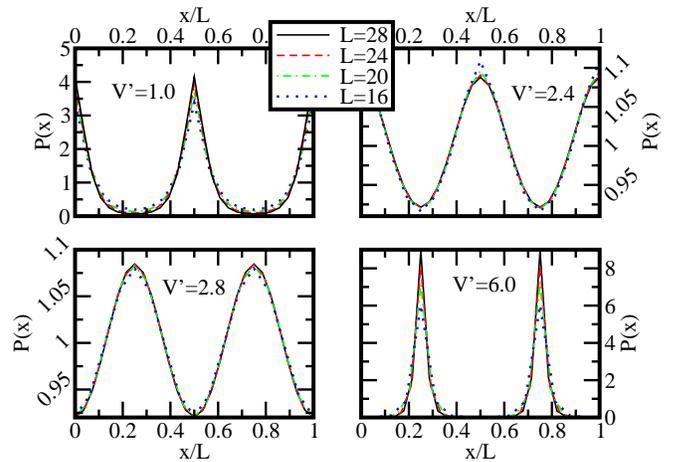}
 \caption{Polarization distributions as a function of the rescaled
   coordinate $x/L$ for systems with $t=1,V=6$ and different values of
   $V'$.  For $V'=1.0$ and $V'=6.0$ (CDW-1 and CDW-2 phases,
   respectively) the polarization distributions exhibit sharp peaks,
   which sharpen with system size.  For the cases $V'=2.4$ and
   $V'=2.8$, on either side of the polarization switch, there are no
   sharp peaks, and the distributions exhibit negligible size
   dependence.  \textcolor{red}{The scaling exponent of the variation
     of the polarization is indicated by arrows in the lower right
     inset of the upper panel of Fig. \ref{fig:pd_tvvp}.}  Exact
   diagonalization calculations with periodic boundary conditions.  }
 \label{fig:finite_tvvp}
\end{figure}

\section{Model Hamiltonian}

\label{sec:Model_method}

The $t-V-V'$ model already has a long
history.~\cite{Hallberg90,Zhuravlev97,Poilblanc97,Mishra11} While
evidence for the four different phases was known since the early
study,~\cite{Hallberg90} the precise phase diagram was only
established recently by Mishra {\it et al.}~\cite{Mishra11} The
Hamiltonian of the $t-V-V'$ model is
\begin{equation}
  \label{eqn:H}
  \hat{H} = \sum_{i=1} \left(- t [ \hat{c}^\dagger_{i+1} \hat{c}_i +
    \mbox{H.c.}] + V \hat{n}_i \hat{n}_{i+1} + V' \hat{n}_i
  \hat{n}_{i+2} \right).
\end{equation}
We take $t$ as the energy scale.  The Hamiltonian can be
mapped~\cite{Hallberg90} onto a spin-$\frac{1}{2}$ chain, via a
Jordan-Wigner transformation,
\begin{equation}
  \label{eqn:H_spin}
  \hat{H} = \sum_{i=1}^L \left(- t [ s_i^+ s_{i+1}^- + s_i^- s_{i+1}^+ ]
  + V s_i^{(z)} s_{i+1}^{(z)}  + V' s_i^{(z)} s_{i+2}^{(z)} \right).
\end{equation}
It is obvious that dihedral $\pi$-rotation symmetry is a symmetry of
the $t-V-V'$ Hamiltonian.  Rotation of each spin by $\pi$ around a
chosen axis, $x$, $y$, or $z$ returns $\hat{H}$ to itself.  Time
reversal and link inversion are also symmetries of $\hat{H}$.  In
Ref. \onlinecite{Pollmann12} these three symmetries were found to be
the ones protecting the topological Haldane phase in odd-$S$ spin
chains.  

The canonical example of the topological Haldane phase is the $S=1$
Heisenberg model.  Crucial insight into the behavior of this model can
be gained via the AKLT variational state whose elementary components
are $S=\frac{1}{2}$ sites, but it is constructed in such a way that
pairs of AKLT sites correspond to a true site of the $S=1$ system.
This construction is also mentioned in Manmana et al.~\cite{Manmana12}
to relate one-dimensional fermion models (or $S=\frac{1}{2}$ models)
to $S=1$ spin models in the example they use, which is an SSH type
model~\cite{Su79} with a Hubbard interaction.

We can proceed in an analogous manner in the $t-V-V'$ model.  We
divide the Hamiltonian in Eq. (\ref{eqn:H_spin}) into two pieces:
\begin{eqnarray}
  \nonumber
  \hat{H}_{\bullet} &=& \sum_{i=1}^{\frac{L}{2}} \left(- t [ s_{2i}^+ s_{2i-1}^- + s_{2i}^- s_{2i-1}^+
  ] + V s_{2i}^{(z)} s_{2i-1}^{(z)}\right) \\
    \nonumber
  \hat{H}_{\bullet \bullet} &=& \sum_{i=1}^{\frac{L}{2}} \left(- t [ s_{2i}^+ s_{2i+1}^- + s_{2i}^- s_{2i+1}^+
  ] + + V s_{2i}^{(z)} s_{2i+1}^{(z)}  \right) \\ 
&&  + V' \sum_{i=1}^{\frac{L}{2}} s_i^{(z)} s_{i+2}^{(z)}.\hspace{.2in}
\end{eqnarray}
Note that $\hat{H}_{\bullet}$ consists of uncoupled pairs of sites.
As in the AKLT procedure, we express $\hat{H}_{\bullet}$ in an $S=1$
(truncated) basis $|+\rangle= |\uparrow\uparrow\rangle, |0\rangle=
\frac{1}{\sqrt{2}} (|\uparrow\downarrow\rangle +
|\downarrow\uparrow\rangle)$, and $|-\rangle=
|\downarrow\downarrow\rangle$.  In this basis $\hat{H}_{\bullet}$
becomes an onsite $S=1$ term,
\begin{equation}
  \hat{H}_{\bullet} = \sum_{i=1}^{\frac{L}{2}} (2 V + t)(S^{(z)}_i)^2 - (t+V)\frac{L}{2}.
\end{equation}
$\hat{H}_{\bullet\bullet}$ turns out to be a pairing term of the form:
\begin{equation}
  \hat{H}_{\bullet\bullet} = \sum_{i=1}^{\frac{L}{2}}
  \left(-\frac{t}{2} [ \hat{S}_{i}^+ \hat{S}_{i+1}^- + \hat{S}_{i}^- \hat{S}_{i+1}^+]
  + (V + 2 V') \hat{S}_{i}^{(z)} \hat{S}_{i+1}^{(z)}\right).
\end{equation}
The Hamiltonian $\hat{H}_{\bullet\bullet}+\hat{H}_{\bullet}$ was
studied~\cite{Oshikawa92} in detail in the context of the Haldane
phase.  The symmetries which protect~\cite{Pollmann12} the Haldane
phase in odd-$S$ systems, namely, time-reversal, dihedral rotation,
and link inversion, are all present in this Hamiltonian, and HAFM and
string order~\cite{denNijs89} are also exhibited.  While this
Hamiltonian is a mapping based on a truncated basis (the $S=0$ states
are missing), below we show that this is not relevant, a Haldane phase
is still exhibited, since the additional terms account for states of
$S_z=0$.

In addition to exact diagonalization of the Hamiltonian
(Eq. (\ref{eqn:H})), we also perform auxiliary calculations.  We
construct a variational wave function, which can be viewed as the
marriage of the AKLT wave function with the Baeriswyl variational wave
function.~\cite{Baeriswyl86,Baeriswyl00} In the AKLT scheme an $S=1$
site is considered as two spin $\frac{1}{2}$ sites.  Bonds connecting
different $S=1$-sites are taken to be in singlet states.  A projector
is applied to the $S=1$-sites themselves, projecting them into
$S_z=1,0,-1$ states.

\begin{figure}[ht]
 \centering
 \includegraphics[width=\linewidth,keepaspectratio=true]{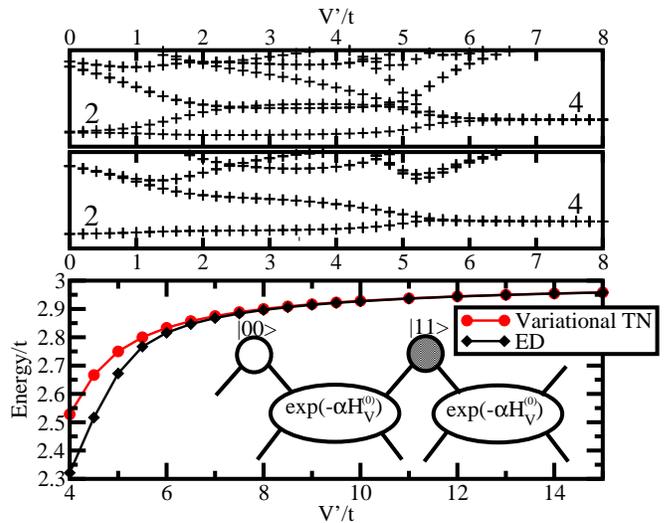}
 \caption{Upper two panels: Low lying states of the entanglement
   spectrum for a system with $V/J=6$ as a function of $V'/J$.  In the
   small CDW-1 phase the degeneracy of the lowest lying state is two,
   while in the CDW-2 phase it is four.  These numbers are indicated
   on the graphs.  The upper of these two graphs is a system with
   $L=24$ with an entanglement cut into a subsystem ($L=12$) and
   environment ($L=12$).  The lower one of the upper two graphs is a
   system with $L=24$ with an entanglement cut into a subsystem
   ($L=11$) and environment ($L=13$).  Exact diagonalization
   calculations with periodic boundary conditions.  Lower panel:
   Comparison of energies for a tensor network variational calculation
   and exact diagonalization.  The variational wave function is drawn
   in the tensor network notation of Ref. \onlinecite{Orus14}.  Both
   calculations use periodic boundary conditions.}
 \label{fig:ESTN}
\end{figure}

In our case, we start with an ordered state for $V' \rightarrow
\infty$, a state with alternating pairs of occupied and unoccupied
sites, $11001100...$, and apply the projector
\begin{equation}
  \exp \left( -\alpha \hat{H}_V^{(0)} \right),
\end{equation}
where $H_V^{(0)}$ is the Hamiltonian without the second-nearest
neighbor coupling term, but including the hopping energy and the
nearest neighbor coupling term.  $\alpha$ is a variational parameter.
The projector is only applied between bonds connecting an occupied and
an unoccupied site ($10$ or $01$).  The wave function is represented
in tensor network notation of Ref. \onlinecite{Orus14} in the lower
panel of Fig. \ref{fig:ESTN}, where a comparison the energies of this
scheme compared to exact diagonalization results is also shown.  The
results indicate that this wave function provides an accurate
description of the system.

In our study of the system with open boundary conditions, we also
complement our exact diagonalization results with a cluster mean-field
theory calculation.  Here clusters of four sites are solved exactly,
but the inter-cluster couplings are considered at the mean-field
level.  We do this for the case of open boundary conditions, meaning
that the mean-field parameters for each cluster vary as a function of
the position of the cluster in the lattice.
  
\section{Polarization amplitude, symmetry analysis, and topological analogy}
\label{sec:polamp}

Topological invariants for non-interacting
systems~\cite{Thouless82,Kane05a,Kane05b,Fu06} are quantities derived
from geometric phases~\cite{Berry84,Zak89} (Zak phase).  The Zak
phase, the topological invariant sensitive to the
transition~\cite{King-Smith93,Resta94} in the SSH model,~\cite{Su79}
has a straightforward many-body generalization~\cite{Resta98,Resta99}
(a single-point Berry phase~\cite{Resta00}).  We define the
polarization amplitude,
\begin{eqnarray}
  \label{eqn:Zq}
  Z_q &=& \langle \Psi | \hat{U}_q | \Psi \rangle, \\ \nonumber
  \hat{U}_q &=& \exp(i 2 \pi q \hat{X} /L).
\end{eqnarray}
where $\hat{X} = \sum_{x=1}^L x \hat{n}_x$.  The operator $\hat{U}_q$ is
known as the total momentum shift operator.  In terms of $Z_q$ the
polarization~\cite{Resta98,Aligia99} of a system with filling $p/q$
per unit cell can be written as
\begin{equation}
  \label{eqn:P}
  P = \frac{L}{2 \pi q} \mbox{Im} \ln Z_q.
\end{equation}
This expression can be shown~\cite{Resta98,King-Smith93} to be
consistent with the modern theory of polarization.  The
variance~\cite{Resta99} of the total position as well as higher order
cumulants~\cite{Souza00} can also be derived~\cite{Hetenyi19}, under
the assumption that $Z_q$ is the analog of a characteristic function,
\textcolor{red}{defined on a discrete set of points ($q$ take only
  integer values).  The polarization of Resta~\cite{Resta98} is the
  first moment of this characteristic function, while the variance of
  Resta and Sorella~\cite{Resta99} is the second cumulant.  Both can
  be obtained~\cite{Hetenyi19} via finite difference derivatives with
  respect to $q$.}  Recently, $Z_q$ was intensively
studied~\cite{Aligia99,Nakamura02,Yahyavi17,Kobayashi18,Hetenyi19,Nakamura19,Furuya19}
as a source of information about quantum phase transitions and the
associated finite size scaling.  In this work, exploiting the fact
that $Z_q$ is a characteristic function, we analyze its Fourier
transform,
\begin{equation}
  P(x) = \sum_{s=0}^{L-1} \exp \left( - i \frac{2 \pi }{L} s x \right) Z_s,
\end{equation}
understood to be the polarization distribution of the system, defined
over the lattice positions $x=1,...,L$.  The summation index $s$ runs
over all components of the polarization amplitude $Z_s$.  The Aligia
and Ortiz~\cite{Aligia99} correction is automatically considered.  For
example, if a system has half filling, $p/q=1/2$, then there will be
no odd-$s$ contributions, and $P(x)$ will have two peaks within one
supercell (see Figs. \ref{fig:pd_tvvp} and \ref{fig:finite_tvvp}).

Due to half-filling, all $Z_q$ for $q$ odd are zero.  In the limiting
cases $V \rightarrow \infty$ and $V'\rightarrow \infty$ $Z_q$ the
nonzero $Z_q$ take the following values~\cite{Hetenyi19}: while
$Z_q=1$ for the former, and $Z_q$ alternates between $\pm 1$ for the
latter.  We can also generally demonstrate the role of link inversion
symmetry by generalizing a result of Zak~\cite{Zak89} to the many-body
case.  In Zak's original paper~\cite{Zak89} it was argued that the Zak
phase takes a trivial value (zero) in the case of inversion symmetry
about a lattice site, while a non-trivial value is taken if the
inversion symmetry is about the bond-center ($\pi$) (also known as
link-inversion symmetry).  Zak showed this by first expressing the Zak
phase,
\begin{equation}
\label{eqn:Zak_k}
\gamma_{Zak} = \frac{i 2 \pi}{a} \int_0^{2 \pi} d k \langle
u_k|\partial_k| u_k \rangle,
\end{equation}
using Wannier functions $w(x)$ as
\begin{equation}
  \label{eqn:Zak_W}
  \gamma_{Zak} = \frac{2 \pi}{a} \int_{-\infty}^{\infty} x |w(x)|^2 dx.
\end{equation}
In Eqs. (\ref{eqn:Zak_k}) and (\ref{eqn:Zak_W}) $a$ denotes the size
of the unit cell and $w(x)$ is the Wannier function for some band.  If
the system obeys reflection symmetry about a lattice site, then $w(-x)
= \pm w(x)$, leading to $\gamma=0$ (equivalent to shifts by $2\pi$).
For the case of link inversion symmetry, $w( - x + a ) = \pm w(x)$,
leads to $\gamma=\pi$.

Our task is to generalize this argument to the many body case.  Our
starting point is the phase of the many-body polarization expression
derived by Resta,~\cite{Resta98,Resta99} applied to a half-filled
system~\cite{Aligia99} (filling $n = p/q$, where $p=1,q=2$),
\begin{equation}
  \label{eqn:Gamma}
  \Gamma = \mbox{Im} \ln Z_2 = \mbox{Im} \ln \langle \Psi|\hat{U}_2| \Psi \rangle.
\end{equation}
We apply a site-centered reflection to the total momentum shift operator,
\begin{equation}
  \hat{R}_s \hat{U}_2 \hat{R}_s^{-1} = \exp(i 4
  \pi \hat{R}_s \hat{X} \hat{R}_s^{-1}/L) = \hat{U}_2.
\end{equation}
In this case the site around which reflection was performed was chosen
to be the one at the origin.  It is easily seen that $\Gamma = -
\Gamma = 0$.  We now apply a reflection operator around a bond center,
using
\begin{equation}
  \hat{R}_s \hat{X} \hat{R}_s^{-1} = \sum_x (L - x + 1) \hat{n}_x,
\end{equation}
leads to the result: $\Gamma = \pi$.  This result was also shown for
matrix product states which are not ``cat states'' (superposition of
two states not connected by any local operator) in
Ref. \onlinecite{Pollmann12}.  Our proof above is entirely general.

In addition to the above result, we can also prove a version of the
LSM theorem relevant to our model.  The original LSM theorem shows
that spin-chains behave qualitatively different, depending on the spin
being integer or half-integer.  In our case, the distinction depends
on whether a model exhibits site or link-inversion symmetry.\\

\begin{figure}[ht]
 \centering
 \includegraphics[width=\linewidth,keepaspectratio=true]{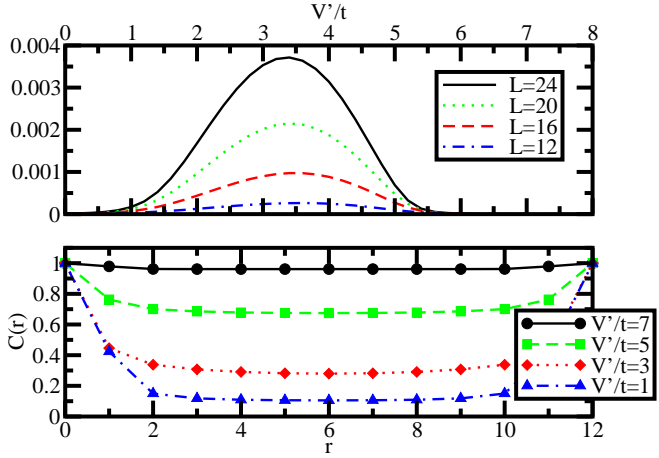}
 \caption{Upper panel: Probability of being in a real-space
   configuration not consistent with $S=1$ hidden antiferromagnetic
   order.  Lower panel: String order correlation function for
   different $V'/t$.  The system size is $L=24$.  Nearest neighboring
   sites were paired.  Exact diagionalization calculations with
   periodic boundary conditions.}
 \label{fig:SOP}
\end{figure}

We start with a ground state $|\Psi_0\rangle$ of a model whose
Hamiltonian consists of a hopping of the type in Eq. (\ref{eqn:H}),
and some coordinate dependent interaction term.  We construct a new
state
\begin{equation}
  |\Psi_1\rangle = \hat{U}_1 |\Psi_0\rangle.
\end{equation}
The energy of this state compared to the ground state is
\begin{equation}
  E_1 - E_0 = -t [ \cos(2 \pi / L) - 1] \sum_i\langle \Psi_0
  |\hat{c}_i^\dagger \hat{c}_{i+1}| \Psi_0 \rangle.
\end{equation}
In the thermodynamic limit, the $E_1 \rightarrow E_0$, but
$|\Psi_1\rangle$ may not be a state that is different from
$|\Psi_0\rangle$.  To show this, apply the different inversion
operators ($\hat{R}_s, \hat{R}_b$) and time reversal symmetry, as
\begin{eqnarray}
  \label{eqn:RsU1Rs}
  \hat{R}_s \hat{T} \hat{U}_1 \hat{T}^{-1} \hat{R}_s^{-1} = & \hat{U}_1   \\ \nonumber
  \hat{R}_b \hat{T} \hat{U}_1 \hat{T}^{-1} \hat{R}_b^{-1} = &-\hat{U}_1.
\end{eqnarray}\\

The state $|\Psi_1\rangle$ is even if the system is site-inversion
symmetric, while odd in the case of bond-inversion symmetry.  The two
possibilities arising from these results are the following.  In the
thermodynamic limit, there may be a gapless excitation which is odd
with respect to bond inversion, alternatively, bond inversion symmetry
may be spontaneously broken with degenerate ground states with a gap
above each.\\

\begin{figure}[ht]
 \centering
 \includegraphics[width=\linewidth,keepaspectratio=true]{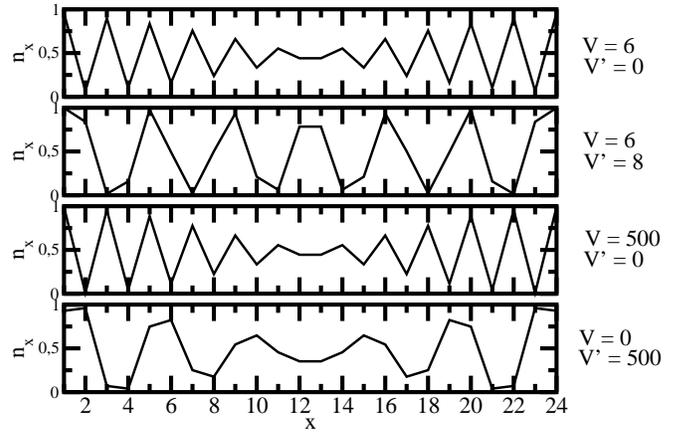}
 \caption{Real space density for four cases, $V=6,V'=0$, $V=6,V'=8$,
   $V=500,V'=0$, and $V=0,V'=500$, under open boundary conditions.
   The system size is $L=24$.}
 \label{fig:n_x}
\end{figure}

One model analyzed by Manmana et al.~\cite{Manmana12} was a spinful
SSH model~\cite{Su79} with a Hubbard interaction.  We show here, that
the results derived above for the $t-V-V'$ model hold for the spinful
SSH model as well.  Let us write it in the following form:
\begin{equation}
  \label{eqn:H_SSHU}
  \hat{H} = \sum_{i,\sigma} \left([- J \hat{c}^\dagger_{i,\sigma} \hat{d}_{i,\sigma} - J' \hat{d}^\dagger_{i,\sigma} \hat{c}_{i+1,\sigma}  + \mbox{H.c.}] +  U \hat{n}_{i,\uparrow} \hat{n}_{i,\downarrow} \right).
\end{equation}
$\hat{c}^\dagger_{i,\sigma}$($\hat{d}^\dagger_{i,\sigma}$) are creation operators
on the different sublattices.  In order to prove either of the above
results, it is sufficient to consider one spin channel, and define a
momentum shift operator of the form
\begin{equation}
  \hat{U} = \exp\left( i \frac{2 \pi}{L} \sum_{j=1}^L j (\hat{n}^{(c)}_{j,\uparrow} + \hat{n}^{(d)}_{j,\uparrow})\right).
\end{equation}
One can use this operator to construct a state $|\Psi_1\rangle = \hat{U}
|\Psi_0\rangle$, and the energy difference will be
\begin{equation}
  E_1 - E_0 = J' [\cos(2 \pi /L) - 1] \sum_i\langle \Psi_0
  |\hat{c}_{i,\uparrow}^\dagger \hat{c}_{i+1,\uparrow}| \Psi_0 \rangle
\end{equation}
Eq. (\ref{eqn:RsU1Rs}) holds meaning that a bond-inversion symmetric
system will have a degenerate ground state.\\

The basis used to construct $\hat{H}_{\bullet}$ and
$\hat{H}_{\bullet\bullet}$ is a truncated one, the state $S=0,M=0$ is
missing.  Still, we can formulate a theorem of the Lieb-Schultz-Mattis
type, as was done above.  The relevant operator is
\begin{equation}
  \hat{U} = \exp\left( i \frac{2\pi}{L} \sum_{j=1}^L j (\hat{S}^z_j + S)\right).
\end{equation}
If one applies the steps above to a fixed $S$ spin model (applying
link-inversion, and time reversal), the result is a sign change in
$\hat{U}$ for an odd-$S$ model, but no sign change for an even-$S$
model.  This is consistent with the results of
Ref. \onlinecite{Pollmann12}.  The fact that the basis is a truncated
one for our case makes no difference, since other states are $S=0$
states, and the maximum spin a site can be is $S=1$.

\section{Exact diagonalization results}
\label{sec:ED}

The results of Mishra {\it et al.}~\cite{Mishra11} for the phase
diagram are shown in Fig. \ref{fig:pd_tvvp}, upper panel.  The phase
lines separate a charge-density wave (CDW-1) a LL, a bond-order and a
second charge-density (CDW-2) wave phase.  Our main result is that in
addition to the known~\cite{Mishra11} phase diagram, we find the
dashed line inside the LL phase which separates phases in which the
polarization (average of the polarization distribution) differs by
one-quarter of a supercell (see Fig. \ref{fig:pd_tvvp} lower panel).
Along the transition line the polarization distribution, $P(x)$ is
flat.  Moving away in either direction gives distributions with maxima
in different places.  Since the filling with respect to number of unit
cells is $1/2$, $\frac{L}{4 \pi } \mbox{Im} \ln Z_2$ corresponds to
the polarization.~\cite{Aligia99} The inset of the upper panel of
Fig. \ref{fig:pd_tvvp} shows the reconstructed distribution $P(x)$ for
selected points along the line $V=6$ with different values of $V'$
below, at and above the transition point (points are indicated in the
main figure with asterisks, upper panel).  The transition in this case
occurs at $V'\approx2.59t$, where $P(x)$ is entirely flat.  The
positions of the two maxima both shift at the transition point.

\begin{figure}[ht]
 \centering
 \includegraphics[width=\linewidth,keepaspectratio=true]{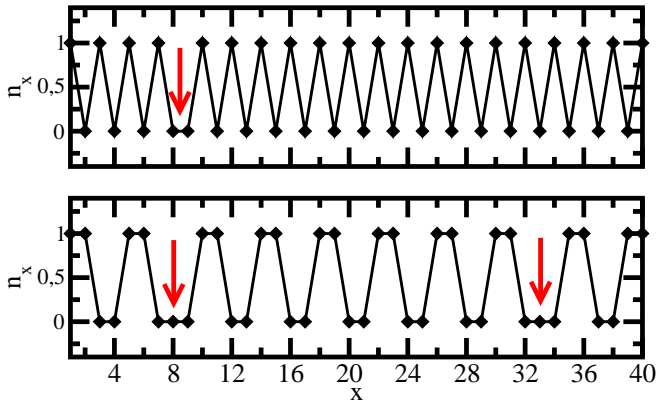}
 \caption{Real space density calculated via cluster-mean-field theory
   for two cases, $V=500,V'=0$, and $V=0,V'=500$, under open boundary
   conditions.  The system size is $L=40$.}
 \label{fig:n_x_cmf}
\end{figure}

Fig. \ref{fig:finite_tvvp} shows the distributions $P(x)$ for
different system sizes for four cases.  The distributions are scaled
by the system size on the $x$-axis to enable comparison.  Two cases
which are in the LL phase are shown ($V'=2.4,2.8$), as well as two
other cases in the two different CDW phases ($V'=1.0$, and $V'=6.0$).
The CDW distributions show sharp peaks, whereas in the LL phase the
distributions have smeared out maxima.  \textcolor{red}{We also
  calculated the size scaling exponent of the variance of the
  polarization, shown in the lower right inset of the upper panel of
  Fig. \ref{fig:pd_tvvp}.  The variance was calculated according to
  the procedure in Ref. \onlinecite{Hetenyi19}.  The size scaling
  exponent was calculated by fitting the variance as a function of
  system size $L$ to the function $f(L)=\alpha L^\gamma$.  Clearly,
  the two LL phases exhibit $\gamma=2$, meaning that the variance of
  the polarization scales as the square of the system size for both
  cases.  This also means that the polarization distributions in the
  $LL$ phase flatten as $L\rightarrow \infty$.  They behave in a
  similar manner to the phase transition line within the LL phase
  shown in Fig. \ref{fig:pd_tvvp}.  Even though finite systems exhibit
  a fixed average polarization, since the variance diverges with
  system size, it will not be an experimentally measurable quantity.
  In contrast, both insulating phases show a $\gamma$ near one.}

\begin{figure}[ht]
 \centering
 \includegraphics[width=\linewidth,keepaspectratio=true]{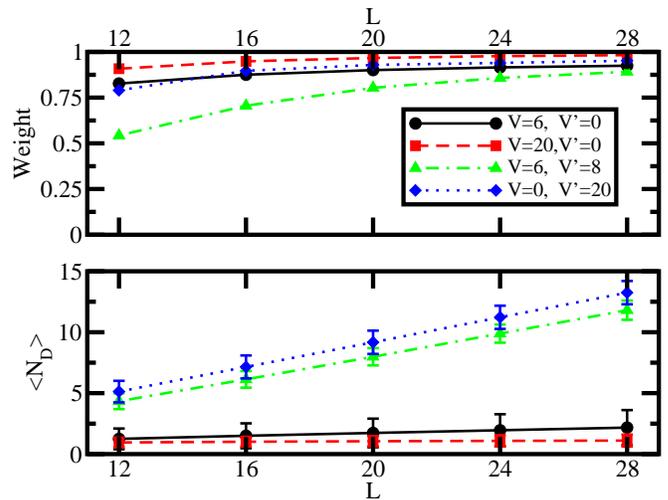}
 \caption{Upper panel: Proportion of configurations in the ground
   state wave-function with number of defects in the ordered state
   with a given parity.  In the CDW-1 cases ($V=6,V'=0$, $V=20,V'=0$,
   both CDW-1) the weight of configurations with an odd number of
   defects is shown, whereas in the CDW-2 cases ($V=6,V'=8$,
   $V=0,V'=20$, both CDW-2) the weight of configurations with an even
   number of defects is shown.  Lower panel: average number of defects
   (appropriate to CDW-1 or CDW-2) and the variance (error bars) in
   the ground state.  The legend applies to both upper and lower
   panels.  Both calculations are exact diagonalization with open
   boundary conditions.  }
 \label{fig:defects}
\end{figure}

The following picture emerges.  The gapless LL phase, in which the
polarization distribution is flat in the limit of large system size,
separates gapped phases in which the polarizations differ by a quarter
of a lattice constant.  This corresponds to a shift in Zak phase of
$\pi$, exactly as in the case of the SSH model.  However, unlike in
that model, the topologically distinct phases are not separated by one
phase transition point.  Instead, they are separated by the entire LL
phase.  In the finite system, the LL phase exhibits a topological
phase line, on either side of which the polarization distribution is
distributed according to a Gaussian whose variance diverges with
system size.  In the insulating phases on either side, the
distributions are peaked and have finite variances, even in the
thermodynamic limit, the polarization therefore is measurable.

The upper panel of Fig. \ref{fig:ESTN} shows two graphs low lying
states of the entanglement spectrum for a system with $V=6$ as a
function of $V'$ for two different entanglement cuts.  The system
consists of $L=24$ sites with periodic boundary conditions.  In one
case (uppermost graph) the entanglement cut is taken at half the
system ($L_s=12$ and the environment is $L_e=12$ which is traced out),
while in the middle panel, the entanglement cut is taken such that the
subsystem size is $L_s=11$, the environment (traced out) is $13$
sites.  In both graphs, the lowest lying state is two-fold degenerate
in the CDW-1 phase, while it is four-fold degenerate in the CDW-2
phase.  In the $L_s=12$ case the two states in the CDW-1 phase form a
basis for a two dimensional irreducible representation of the link
inversion operator.  The four states in the CDW-2 phase form a basis
for two two dimensional irreducible representations of the link
inversion operator.  For the $L_s=11$ case the two degenerate states
on the CDW-1 side form two identity representations of the
site-inversion operator, while on the CDW-2 side, we find, again, two
two-dimensional representations thereof.

We also calculated the entanglement spectrum (for a system with $L=24$
and an entanglement cut at half the system) for the variational ansatz
shown in Fig. \ref{fig:ESTN}.  Since the symmetry is explicitly broken
in this variational wave function, there are four different ways to
choose such an entanglement cut.  Each different cut gives rise to one
low-lying state in the entanglement spectrum, consistent with the
fourfold degneracy of the CDW-2 phase found via exact
diagonalization.

In Fig. \ref{fig:SOP} we show quantities related to HAFM ordering.  By
hidden antiferromagnetic (AFM) order we mean~\cite{Oshikawa92} the
following.  Once sites are paired, as was done to construct the
truncated Hamiltonian in section \ref{sec:Model_method}, the
$z$-component of the spin of the paired sites are calculated (if both
sites are occupied, $S_z = +$, if one site of the two is occupied,
$S_z=0$, if both sites are empty, $S_z=-$).  In hidden AFM order $+$
and $-$ sites must alternate, possibly with $0$ sites in between.  An
example of such a configuration is $+000-+-00+0-...$ (a configuration
in the space of paired sites).  In the upper panel of
Fig. \ref{fig:SOP} the fraction of configurations {\it not} consistent
with hidden AFM order are shown.  We see that in the CDW-1 and CDW-2
states, the overwhelming majority of configurations are consistent
with hidden AFM.  It is only in the ungapped region (mainly the LL
phase) where configurations not consistent with hidden AFM are found.

The lower panel of Fig. \ref{fig:SOP} shows the string order
correlation function of den Nijs and Rommelse~\cite{denNijs89}, which
is of the form
\begin{equation}
    C(r=k-j) = \langle S_j^z \exp \left( i \pi \sum_{l=j+1}^{k-1} S_l^z \right) S_k^z \rangle.
\end{equation}
The CDW-1 state displays a rapidly decaying correlation function,
while the CDW-2 state show ordering.  For the former, definite
conclusions are difficult to draw, due to the small system size, but
it appears that there is a decay in the string correlation function. 

Manmana et al.~\cite{Manmana12} analyzed the BBCP in an interacting
topological system, a spin-dependent SSH model with Hubbard on-site
interaction.  This model is presented as an example for symmetry
breaking at the edges, as opposed to single-particle topological edge
states corresponding to poles or zeros of the single-particle Greens's
function.  Symmetry breaking at the edges is a many-body phenomenon to
which the single-particle Green's function based topological invariant
is not necessarily sensitive.

Their analysis is not directly applicable to our model due to the
finite range of the interactions.  We are not able to take a dimerized
limit.  We can, however, investigate a system with open boundary
conditions.  The density distribution for our interacting system with
open boundary conditions is shown for four cases in Fig. \ref{fig:n_x}
under open boundary conditions.  The upper two plots show a CDW-1 ($V
= 6, V' = 0$) and CDW-2 ($V = 6, V' = 8$) examples, while the bottom
two show nearly completely ordered CDW-1 and CDW-2 cases.  The
completely ordered states are interesting because each one is
adiabatically connected to all states on the same side of the LL
region.  All distributions in Fig. \ref{fig:n_x} invert around the
midpoint, meaning that the left edge can be related to the right edge
via link inversion symmetry.  In Fig. \ref{fig:n_x_cmf} the density
distributions are shown for the nearly completely ordered CDW-1 and
CDW-2 states based on a four-site cluster mean-field calculation with
open boundary conditions.  In each case one particular symmetry broken
state exists at each edge.  In the middle the system has to
``connect'' between the different symmetry broken states at the edges.
For the CDW-1 case one (or an odd number) of defect(s) is needed,
while in the CDW-2 case two (or an even number of) defect(s) are
required.  The defects are indicated by the arrows on the figure.

Defects can be located anywhere on the lattice, and the quantum ground
state can be a superposition of states with different numbers of
defects in different places.  Fig. \ref{fig:defects} shows the weight
of configurations with an odd(even) number of defects for the
CDW-1(CDW-2) cases in the ground state wave function.
\textcolor{red}{In these calculations, the ground state wave function
  was obtained in real space.  In each real-space component the
  defects were counted as follows.  In the CDW-1 case, there are two
  ordered states, $1010...$ or $0101...$, where $0$ represents an
  empty site, $1$, an occupied one.  If the left most site is filled,
  we assume that segment of the system is in the former, if not, then
  the latter.  We then check, starting from the leftmost site, going
  right, whether the configuration deviates from this ordered state.
  For example, if the configuration is $10100...$, the fifth site
  exhibits a defect.  After a defect is encountered, we reset the
  ordered state accordingly, and look for the next defect.  We use a
  similar scheme for the CDW-2 state, except there, the possible
  ordered states are $11001100...$, $01100110...$, $10011001...$, and
  $00110011...$.  This means that the first two sites determine an
  ordered state.  When a defect is encountered the resetting to a new
  ordered state is based on the defect site, and the one before it.}
The defects counted are of the types indicated in
Fig. \ref{fig:n_x_cmf}.  As the system size increases, the weight of
configurations with odd(even) number of defects approaches unity in
the CDW-1(CDW-2) ground state.  The lower panel of
Fig. \ref{fig:defects} shows the average number of defects and its
variance for the four cases.  The number of defects increases linearly
with system size.

As mentioned above, it is difficult to construct a model in this case,
which identifies the edge state.  In the example of Manmana et
al.~\cite{Manmana12} the model was one in which dimerization was
possible, and in this limit, the analysis is not much more difficult
than for the non-interacting SSH model.  In our case, for the CDW-2,
such a limit does not exist, however, we can make conjectures based on
the results above.  It appears that symmetry breaking occurs at the
edges, and in order to connect the two boundaries, an odd number of
defects is necessary in the CDW-1 phase, and the number of defects
needs to be even in the CDW-2 phase.  This may be a common scenario in
ordinary symmetry broken systems, but most importantly, it coincides
with the findings of Manmana et al.~\cite{Manmana12} In the case of
symmetry breaking at the edges in an topological interacting system, a
localized edge state may not be easily identifiable, since it is a
many-body state, not a single-particle one.

\section{Conclusion}
\label{sec:con}

The main criterion for topological insulation is
the topological invariant assuming non-trivial values.  In
non-interacting systems the invariant undergoes a finite change at a
gap closure point.  Topologically distinct phases are separated by gap
closure points, or quantum phase transitions.

In the $t-V-V'$ model, for fixed $V$ large enough to start in a
charge-density wave phase, as $V'$ is increased a Luttinger liquid
phase is encountered.  Passing through this phase, there is a bond
order phase followed by a new charge density wave.  The many-body
polarization single-point Berry phase changes discontinuously inside
the Luttinger liquid phase at a critical $V'$.  However, in contrast
to non-interacting systems, this change in the topological invariant
occurs inside a gapless phase.  Inside the gapless Luttinger liquid
phase the variance of the polarization diverges with system size,
meaning that the Berry phase is undefined inside this whole region.
In short, topologically distinct phases are separated, not by a
quantum phase transition point, as in non-interacting systems, but by
the Luttinger liquid phase itself.  The topological phase was shown to
be a Haldane phase, exhibiting hidden anti-ferromagnetic order and
finite string correlation.

The model we studied can be realized experimentally in a cold atoms in
optical lattice setting.~\cite{Bloch08} The degree of control in such
experiment places strongly correlated one-dimensional models within
reach.~\cite{Cazalilla11} Particularly pertinent to our study is that
the geometric phase~\cite{Berry84,Zak89} which gauges the transition
can also be directly~\cite{Atala13} measured.

\section*{Acknowledgments} This research was supported by the National
Research, Development and Innovation Fund of Hungary within the
Quantum Technology National Excellence Program (Project Nr.
2017-1.2.1-NKP-2017-00001).

\end{document}